\documentstyle[12pt]{article}
\def\bib{\bibitem}
\def\be{\begin{equation}}
\def\ee{\end{equation}}
\def\barr{\begin{array}}
\def\earr{\end{array}}
\def\ie{ {\it i.e.}}
\def\eg{ {\it e.g.} }
\def\etc{ {\it etc.} }
\def\etal{ {\it et al.} }
\def\lsim{\:\raisebox{-0.5ex}{$\stackrel{\textstyle<}{\sim}$}\:}
\def\gsim{\:\raisebox{-0.5ex}{$\stackrel{\textstyle>}{\sim}$}\:}
\def\rp{$R_p \hspace{-1em}/\;\:$}
\def\rMS{$R_p$MSSM}
\def\rpMS{$R_p \hspace{-1em}/\;\:$MSSM}
\def\gev{\: {\rm GeV} }
\def\pb{\: {\rm pb}}
\def\ra{\rightarrow}
\def\Ch{{\chi}_1}
\def\Ne{{\chi^0_1}}

\def\ib#1,#2,#3{       {\it ibid.\/ }{\bf #1} (19#2) #3}
\def\ap#1,#2,#3{       {\it Ann.~Phys.~(NY)\/ }{\bf #1} (19#2) #3}
\def\ijmp#1,#2,#3{     {\it Int.~J.~Mod.~Phys.\/ } {\bf A#1} (19#2) #3}
\def\mpl#1,#2,#3 {     {\it Mod.~Phys.~Lett.\/ } {\bf A#1} (19#2) #3}
\def\np#1,#2,#3{       {\it Nucl.~Phys.\/ }{\bf B#1} (19#2) #3}
\def\npps#1,#2,#3{     {\it Nucl.~Phys.~B (Proc.~Suppl.)\/ }{\bf B#1}
                             (19#2) #3}
\def\plb#1,#2,#3{      {\it Phys.~Lett.\/ }{\bf B#1} (19#2) #3}
\def\pr#1,#2,#3{       {\it Phys.~Rev.\/ }{\bf D#1} (19#2) #3}
\def\prep#1,#2,#3{     {\it Phys.~Rep.\/ }{\bf #1} (19#2) #3}
\def\prl#1,#2,#3{      {\it Phys.~Rev.~Lett.\/ }{\bf #1} (19#2) #3}
\def\pro#1,#2,#3{      {\it Prog.~Theor.~Phys.\/ }{\bf #1} (19#2) #3}
\def\rmp#1,#2,#3{      {\it Rev.~Mod.~Phys.\/ }{\bf #1} (19#2) #3}
\def\sp#1,#2,#3{       {\it Sov.~Phys.-Usp.\/ }{\bf #1} (19#2) #3}
\def\zpc#1,#2,#3{      {\it Zeit.~f\"ur Physik\/ }{\bf C#1} (19#2) #3}

\begin{document}
\setcounter{page}{0}
\renewcommand{\thefootnote}{\fnsymbol{footnote}}
\thispagestyle{empty}
\vspace*{-1in}
\begin{flushright}
SCIPP-96/27\\[1.7ex]
MPI-PTh/96-44\\[1.7ex]
{\large \tt hep-ph/9606415} \\
\end{flushright}

\vskip 45pt
\begin{center}
{\Large \bf ALEPH Four-Jet Excess, $R_b$ and $R-$Parity 
Violation}\footnote{Work supported in part by the Polish Committe for 
                    Scientific Research and the European Union grant
                    "Flavourdynamics"  CIPD-CT94-0034.}

\vspace{11mm}
\bf
 Piotr H. Chankowski$^{(1,2),}$\footnote{chank@fuw.edu.pl},
 Debajyoti Choudhury$^{(3),}$\footnote{debchou@mppmu.mpg.de}
 {\rm and} Stefan Pokorski$^{(2,3),}$\footnote{stp@mppmu.mpg.de}

\rm
\vspace{13pt}
$^{(1)}$ {\em Santa Cruz Institute for Particle Physics, University
of California,\\ 
Santa Cruz, CA 95064, U.S.A.} \\[2ex]

$^{(2)}$ {\em Institute of Theoretical Physics, Warsaw University,\\
Ho\.za 69, 00-681 Warsaw, Poland} \\[2ex] 

$^{(3)}${\em Max--Planck--Institut f\"ur Physik,
              Werner--Heisenberg--Institut,\\
              F\"ohringer Ring 6, 80805 M\"unchen,  Germany.}

\vspace{50pt}
{\bf ABSTRACT}
\end{center}

\begin{quotation}
We review briefly the indications for some relatively light 
superpartners based on the ~$R_b$ ~anomaly and discuss the 
dependence of the potential increase in $R_b$ 
on the assumption about ~$R-$parity (non)conservation.
We point out that the exotic 4-jet events reported by
ALEPH may constitute a signal for supersymmetry with such a light
spectrum and with explicitly
broken $R-$parity. A parton level simulation shows that production 
of a pair of light charginos with their subsequent baryon-number
violating decays (either through a stop or through
a neutralino) could possibly give rise to this excess. The decay 
$\Ch^- \ra \tilde{t}_1 b \ra d s b$ 
with $m_{\chi^-_1} \sim 60 \gev$ 
and $m_{\tilde t} \sim 52 \gev$ leads to signatures very close 
to the experimental observations.
\end{quotation}

\vfill
\newpage
\setcounter{footnote}{0}
\renewcommand{\thefootnote}{\arabic{footnote}}

\setcounter{page}{1}
\pagestyle{plain}
\advance \parskip by 10pt

Preliminary results from the LEP1.5 run (after the upgrade of energy to ~
$\sqrt{s}=130-136$ GeV) ~include peculiar four-jet events reported by
ALEPH \cite{ALEPH4J}. They join the other, longstanding, ``LEP puzzles''---
the ~$R_b$ ~and ~$R_c$ ~anomalies. The presently measured values ~
$R_b=0.2211\pm0.0016$ ~and ~$R_c=0.1598\pm0.0069$ \cite{LEPEWWG} are ~
$3.4\sigma$ ~and ~$1.7\sigma$ ~away from their Standard Model (SM) 
values (for ~$m_t=175$ GeV). The ~$R_b$ ~anomaly is statistically 
more significant than the 
$R_c$ ~one and it is reasonable to consider theoretical scenarios which
predict a ~$R_b$ larger than that in the SM, but with ~$R_c$ the same.
Even if ~$R_c$ ~is fixed to its SM value ~$R_c=0.172$, ~the value
of ~$R_b$ ~measured under this assumption, ~$R_b=0.2202\pm0.0016$, ~is
still ~$\sim3\sigma$ ~away. Considerable excitement has been caused by
realization that the MSSM ({\em viz.} the minimal supersymmetric extension 
of the SM) presents such a scenario
provided some of the new particles are light \cite{MY_MSSM,KANE,ELONA}. 
Specifically, it has been shown that significant increase in ~$R_b$ ~is
possible for small ~$\tan\beta$ ~with light chargino and stop, and for
large   ~$\tan\beta$ ~with light ~$CP-$odd Higgs boson and chargino and stop.
These conclusions have been reached in the framework of 
$R-$parity conserving MSSM (henceforth called \rMS).

In this note, we extend the discussion of ~$R_b$ ~to the case of MSSM 
with ~$R$-parity broken explicitly (\rpMS) and show that a stronger 
increase in $R_b$ can be expected and for a larger region of the 
parameter space (in particular, the relevant range for ~$\tan\beta$ ~
is wider). We then 
speculate on the physical significance of the ALEPH events and 
point out that they can be interpreted as having arisen from a
chargino pair decaying {\em via} $R$-parity violating interactions. 
While both stop-- and neutralino--mediated decays are possible, 
the former is distinctly the better alternative, especially if 
the stop is lighter than the chargino. 
Signatures closest to the experimental one are obtained for ~
$m_{\chi^+_1}\sim 60$ GeV ~and ~$m_{\tilde t_R}\sim 55$ GeV. ~
Such an interpretation also 
leads to a significant additional contribution to ~$R_b$.

We begin by briefly discussing the results for ~$R_b$. ~Within ~
$R-$parity conserving MSSM\footnote{We assume for this case gaugino mass 
                                    unification, {\em viz.} ~
                                    $M_1 = (5/3) \tan^2\theta_W M_2$.},
the increase of ~$R_b$ ~is not unconstrained. Not only must
the perfect fit in the SM to the bulk of the precision data be
maintained, several other experimental constraints must be
satisfied (see \cite{MY_RB,ELONA} for an extensive discussion) too. 
It has been shown that from this global point of view, in a realistic 
\rMS, ~$R_b$ ~can  be as large as ~0.2180(0.2185) ~for low and 
large ~$\tan\beta$ ~respectively. Although still ~$1.5(1.0)\sigma$ ~away
from ~$0.2202$ ~this is an interesting improvement over the SM 
prediction. In the first place, the overall ~$\chi^2$ ~is much better.
Moreover, the fitted value of ~$\alpha_s(M_Z)$ ~is lower than in the SM
fits to precision data ~($0.116\pm0.005$ ~and $0.122\pm0.005$ respectively)
thus affording better agreement with the values obtained from scaling
violation in the deep inelastic scattering or lattice 
calculations\footnote{The overall average of ~$\alpha_s(M_Z)$ ~from the 
                      deep inelastic scattering, ~$\tau$ ~decays and jet 
                      physics is $0.117\pm0.003$ ~\cite{BETHKE}.}.

If ~$R-$parity is broken, several constraints coming from direct sparticle
searches are relaxed.  In particular, the D0 exclusion plots~\cite{D0_EXCL}
in the ~$(m_{\tilde t}, m_{\Ne})$ ~plane are no longer valid~\cite{DPROY}
and limits on the ~$(\mu, M_2)$ ~plane from chargino searches at LEP1.5 ~
($m_{\chi^+}>65$ GeV) ~and neutralino serches at LEP1 and LEP1.5 
are relaxed. However, for small gaugino masses, the decays ~
$Z^0\ra \chi^0_i \chi^0_j$ ~can significantly contribute to the total ~
$Z^0$ ~width, thus worsening the ~$\chi^2$ ~fit for small chargino masses.
This effect is almost independent of the assumptions regarding ~$R-$parity,
but can be circumvented by relaxing the assumption of gaugino mass 
unification. 
The quality of the fit to the remaining observables depends on 
parameters irrelevant for ~$R_b$, and is not in conflict
with an increase in the latter.

A sample of results for ~$R_b$, as a function of the chargino mass, 
obtained from a global fit in the MSSM to the precision electroweak data,
without imposing constraints relaxed by ~$R-$parity violation and 
without assuming gaugino mass unification (\ie neglecting the contribution
of ~$Z^0\ra\chi^0_i\chi^0_j$ ~to the total ~$Z^0$ ~width) are shown in Figs. 1
and 2 together with the corresponding values of the overall ~$\chi^2$. ~
(The analogous fit in the SM gives ~$\chi^2\approx18.4$.) 
The particular dependence of ~$R_b$ ~on the chargino mass arises
due to specific properties of the chargino mass matrix 
which are strongly asymmetric with respect to the change of sign of the ~
$\mu$ ~parameter. 

\begin{figure}[ht]
\vskip 6.0in\relax\noindent\hskip -0.7in\relax{\includegraphics{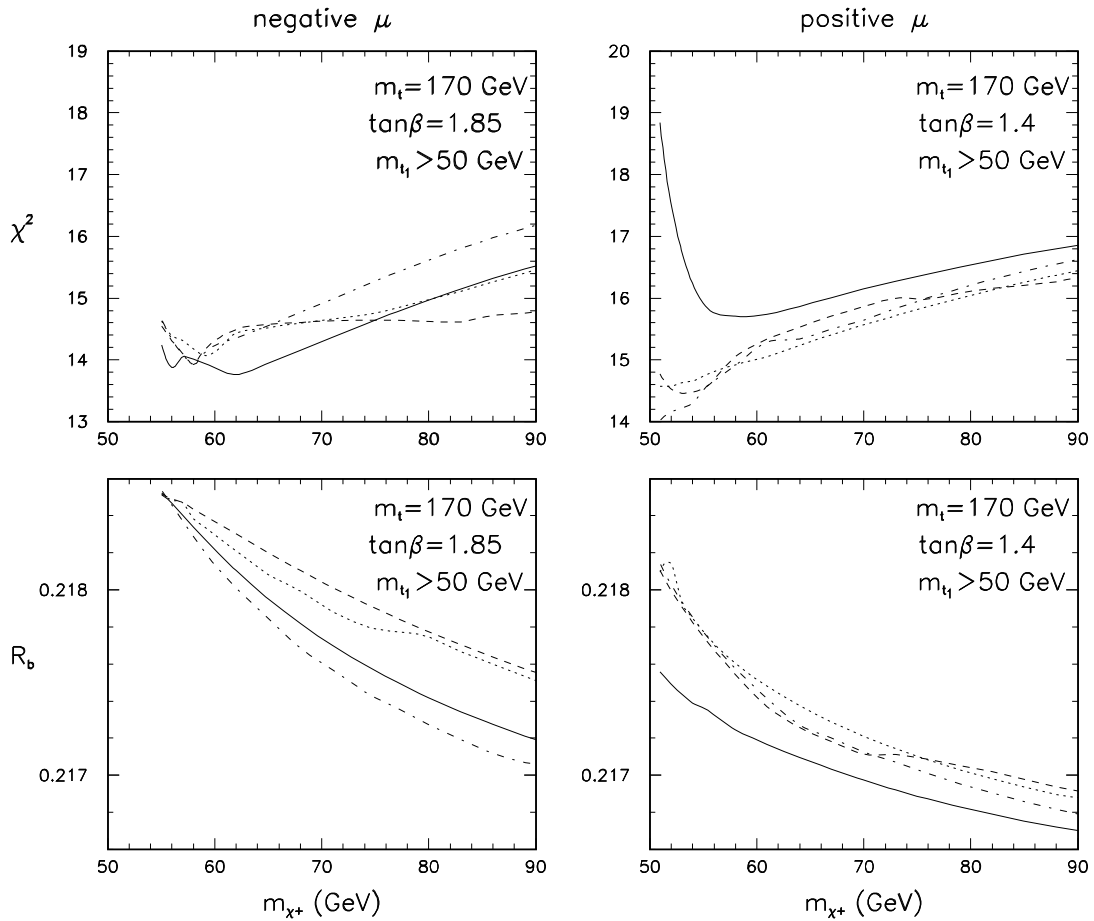}}
\vspace{-18ex}
\caption{ {\em $\chi^2$ (upper pannels) as a function of $m_{\chi^+}$ for 
$r\equiv M_2/|\mu|=0.5$ (solid lines), $1$ (dashed), $1.5$ (dotted)
and $3$ (dash-dotted) for $m_t=170$ GeV, $\tan\beta=1.85$ for $\mu<0$,
$\tan\beta=1.4$ for $\mu>0$ and
$M_A=m_{\tilde t_2}=1$ TeV. In lower panels the best 
values of $R_b$ with the restriction $\chi^2< \chi^2_{min} +1$ (here 
$\chi^2_{min}$ denotes the best $\chi^2$ for fixed value of $m_{\chi^+}$) 
are shown.} }
\label{fig:rb_1}
\end{figure}

We see from Fig.\ref{fig:rb_1} that, for low ~$\tan\beta$ ~and ~$\mu>0$, ~
$R_b$ ~(the overall ~$\chi^2$) ~has a maximum (minimum) around ~
$R_b\approx0.2180$ ~
for light charginos ~($\sim50$ GeV). For heavier chargino masses, ~$R_b$ ~
decreases rather rapidly (\eg for ~$m_{\chi^+}>60$ GeV ~the best ~$R_b$ ~
is already smaller than ~0.2176). ~
For ~$\mu<0$, a stronger enhancement of ~$R_b$ ~is possible. As explained in
ref. \cite{MY_RB}, this occurs for charginos that are strong 
gaugino--higgsino mixed states ~($r\equiv M_2/|\mu|\sim0.5-1.5$), ~\ie\ with 
both ~$\bar{b}\tilde{t_R}\chi^-$ ~and ~
$Z^0\chi^+\chi^-$ ~couplings enhanced.
This branch can support chargino masses in the 50--60 GeV range, but only for
a larger ~$\tan\beta$ ~($\gsim 1.8$) ~as compared to the other branch. 
Nevertheless, the values of ~$R_b$ ~for the same chargino and stop 
masses are larger than for positive ~$\mu$. ~Although an 
increase in ~$\tan\beta$ ~leads to a decrease of the top Yukawa coupling 
(and hence in the ~$\bar{b}\tilde{t_R}\chi^-$ ~vertex), this is 
more than compensated by the increase in the ~$Z^0\chi^+\chi^-$ ~vertex.
Thus, for ~$m_{\Ch}\approx55\gev$ ~and ~$m_{\tilde t}\approx50 \gev$, ~one 
obtains ~$R_b\approx 0.2185$. ~With \rp  ~one can take full advantage of 
the mechanism detailed in ref. \cite{MY_RB}, which now operates for very 
light charginos. It is interesting to observe that in the \rpMS\ a sizable 
increase in ~$R_b$ ~is possible for a relatively large ~$\tan\beta$ ~range ~
$(1-2)$, ~with low ~$\tan\beta$ ~values for ~$\mu >0$, ~and 
larger values for ~$\mu < 0$.

\begin{figure}[ht]
\vskip 5.0in\relax\noindent\hskip -0.8in\relax{\includegraphics{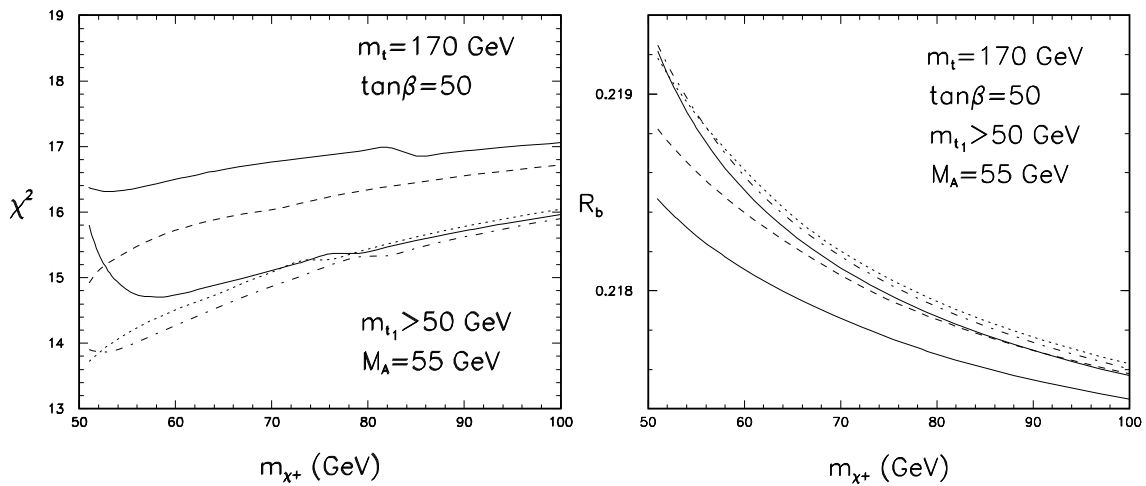}}
\vspace{-30ex}
\caption{ {\em $\chi^2$ (left pannel) as a function of $m_{\chi^+}$ ($\mu>0$) 
for $r\equiv M_2/|\mu|=1$ (upper solid lines), $1.5$ (dashed), $3$ (dotted), 
$5$ (dash-dotted) and $10$ (lower solid) for $m_t=170$ GeV, $\tan\beta=50$.
$m_{\tilde t_2}=1$ TeV, $M_A=55$ GeV, 
$m_{\tilde b_R}=130$ GeV.  In the right panel the best 
values of $R_b$ (with the restriction $\chi^2< \chi^2_{min} +1$ as in Fig.1)
are shown.} }
\label{fig:rb_2}
\end{figure}
Significant enhancement of ~$R_b$ ~is also possible for large ~$\tan\beta$ ~
values, ~$\tan\beta\approx m_t/m_b$. ~In this case, in addition to the
stop--chargino contribution
there can be even larger positive contribution from
the ~$h^0$, ~$H^0$ ~and ~$A^0$ ~exchanges in the loops, provided those
particles are sufficiently light (in this range of ~$\tan\beta$, ~
$M_h\approx M_A$) ~and non-negligible sbottom-neutralino loop
contributions\footnote{The claim of ref. \protect\cite{KANE} that such a 
                       scenario is ruled out, has not yet been confirmed by 
                       an experimental analysis.}

In Fig.\ref{fig:rb_2},
we show the dependence of ~$\chi^2$ ~and ~$R_b$ ~on the chargino mass for a
particular $(m_t, \tan\beta)$~ combination. As in Fig.\ref{fig:rb_1},
we have imposed the condition ~$m_{\tilde t_1}>50$ GeV. ~
The sbottom mass was fixed to ~130 GeV. ~The main
difference with the low ~$\tan\beta$ ~case is the independence of the
results on the sign of ~$\mu$ ~(which can be traced back to the approximate
symmetry of the chargino masses and mixings under $\mu \ra -\mu$)
and monotonic decrease of ~$R_b$ ~with increasing chargino mass.
However, due to the combined effect of ~$A^0$ ~and chargino-stop and
neutralino-sbottom exchanges, ~$R_b$ ~remains as large as ~$0.2178$ ~even
when all three masses ~$M_{A^0}$, ~$m_{\chi^+_1}$ ~and ~$m_{\tilde t_1}$ ~are
taken to be ~65 GeV. 

Thus, in \rpMS, $R_b$ can be much larger than is possible in \rMS.
The main change is that, with ~$R-$parity broken, the 
phenomenologically acceptable range of chargino masses 
is much wider than with ~$R-$parity conserved. In particular, for low ~
$\tan\beta$ ~both regions of positive and negative ~$\mu$ ~are 
still of interest, and ~$R_b\sim0.2180 (0.2185)$ ~is 
still possible for positive (negative) ~$\mu$.

In the context of \rpMS\ one may speculate on
the physical significance of the ALEPH four-jet events (apart from the light
gluino window \cite{FARRAR}, lack of 
missing energy in these events excludes any 
interpretation within the \rMS).
One may even ask if both ~$R_b$ ~and ALEPH events can have simultaneous
explanation.

$R-$parity is defined as $R_p = (-1)^{3B+L+2 S}$ \cite{RPARDEF}
(with $B,L,S$ referring to the baryon number, lepton number
and the intrinsic spin respectively). It serves to eliminate
certain $F-$terms that are otherwise allowed within the
MSSM~\cite{RPAR}. With $L_i$, $E^c_i$, $Q_i$, $U^c_i$ and
$D^c_i$ denoting the superfields, these extra pieces in the
superpotential can be parametrized as
\be
W = \frac{1}{2} \lambda_{ijk} L_i L_j E^c_k
  + \lambda'_{ijk} L_i Q_j D^c_k
  + \frac{1}{2} \lambda''_{ijk} U^c_i D^c_j D^c_k
     \label{superpot},
\ee
where $\lambda_{ijk} = -\lambda_{jik}$ and
$\lambda^{\prime\prime}_{ijk} = -\lambda^{\prime\prime}_{ikj}$.
While the first two terms in eq.(\ref{superpot}) violate
lepton number ($L$), the last one violates baryon number ($B$), and
simultaneous presence of both can lead to rapid proton decay.
Although $R_p$ is a sufficient condition for suppressing proton
decay, it is not a necessary one. In fact, the imposition of
either baryon number or lepton number conservation is quite
sufficient\footnote{Even this might be too strong a
          requirement \protect\cite{CRS}.}. Investigation of the
signature and possible consequences of a \rp\
scenario is thus an worthwhile activity~\cite{RPAR2}.

The ~$L$-violating couplings in eq.(\ref{superpot}) are constrained
mainly from data on lepton and meson decays \cite{BGH}. Additional
constraints also come from the lack of observation of a Majorana
mass for the ~$\nu_e$ ~\cite{MAJORANA} as well as LEP data on ~$Z^0$ ~
partial widths \cite{BES}. In the absence of ~$L-$violation, the
strongest constraints on the ~$\lambda^{\prime\prime}$ ~couplings
come from low energy ~$\Delta B = 2$ ~processes such as ~$n-\bar n$ ~
oscillations or double nucleon decays \cite{GOSH}. Further constraints
can also be derived from the requirement
that Yukawa couplings remain perturbative until the GUT
scale~\cite{BR} or from hadronic decay widths of the 
$Z^0$~\cite{BCS}. However, most such constraints turn out to be much
weaker than the typical bounds on the $L-$violating couplings.
Yet, most of the search strategies \cite{LEP2,BKP} have focussed on the
latter type. Part of the reason can be ascribed to the perception
that ~$B-$violation at the SUSY scale can wash out GUT scale
baryogenesis \cite{COSMO}. However, Dreiner and Ross \cite{HDR} have
pointed out that such bounds are model dependent and can easily
be evaded. 

It is important to comment on
the direct effect of ~\rp\ couplings on the global fits and ~$R_b$ 
in particular.
As the analyses of refs.\cite{BES,BCS} show, for moderate values
of these couplings (of the order of the Yukawa couplings), 
these extra contributions to the electroweak precision variables are small. 
We have checked that the contribution to $R_b$ remains at the level 
$10^{-4}$ for a large range of parameters. Qualitatively, it follows
from the fact that any loop contribution with 
$\lambda^{\prime\prime}$--type coupling is suppressed by the fourth 
power of ~$\sin\theta_W$. ~

Moreover, unless the relevant couplings are surprisingly large, 
implementation of ~\rp\  ~into the MSSM has little impact on
the cross sections for the production of superpartners.
The decay signatures are drastically changed though. Since the
lightest supersymmetric partner (LSP) is no longer stable, it will,
normally, decay within the detector thus calling for a modification
of most search strategies. One may then
speculate that the recently observed 
$4-$jet events in the ALEPH detector at LEP1.5 originate in the production
of a supersymmetric particle which subsequently decays via \rp\ couplings.

With an accumulated
luminosity of $\sim 5.7 \pb^{-1}$ at $\sqrt{s} = 130 - 136$ GeV, and after
the application of several selection criteria, ALEPH
observes 16 events with 4 (or 5) jets where the SM expectation is
only 8.6. Of these events, 5 are peaked around a combined mass of
105 GeV, with a spread of 6.3 GeV. Further, the events are consistent
with no missing energy. As the SM expectation for the
same interval is only 0.4 events, the probability that the excess can be
explained by fluctuations is less than $10^{-4}$.
Assuming the excess events to be a signal,
the data is indicative \cite{ALEPH4J} of a particle ($X$) of mass
$m_X\sim55$ GeV with a pair-production cross section of
\be
\sigma(e^+e^-\ra X\bar{X}; \sqrt{s}\sim135\gev)=(3.7\pm1.7)\pb,
\label{aleph-cs}
\ee
Of particular significance is the
fact that {\em at best 1 event} is consistent with a ~$b\bar{b}b\bar{b}$ ~
final state, thus ruling out the possibility of the supersymmetric Higgs 
boson production, ~$Z^0\rightarrow A^0h^0$. ~
This very fact renders difficult any explanation within the \rMS.
The only suggested counterexample \cite{FARRAR}
interprets these events to be pair-produced squarks decaying into a
quark and a {\em light} gluino ~($m_{\tilde g}\sim1\gev$). ~As the latter
decays only after hadronization, the (missing) energy carried by the 
daughter photino is too small to be a good signature. While interesting,
this scenario is not particularly helpful to the explanation
of the ``$R_b$ ~anomaly''.

Within the regime of ~\rpMS, the simplest explanation would then be one 
where ~$X$ ~is one of the sfermions and is also the lightest supersymmetric
particle
\footnote{Since the LSP is no longer stable in the 
          presence of ~\rp, ~one is not constrained, 
          phenomenologically, to require it to be a neutralino.}.
It would then decay via  the couplings of eq.(\ref{superpot}) into
two fermions. Since an explanation of the ~$R_b$ ~anomaly prefers a
relatively light ~$\tilde{t}_R$, ~it is tempting to identify it with ~$X$. ~
The low production cross section ~($\sim 0.4$ pb) ~rules out such an
explanation though. In fact, only two kinds of sfermions, ~$\tilde{\nu}_e$ ~
and ~$\tilde{d}_{iL}$ ~may have the requisite large
production cross section. The cross-section for ~$\tilde{\nu}_e$ ~production
may be enhanced in the presence of a relatively light (and gaugino-dominated)
chargino \cite{BKP}. With the sneutrino decaying through 
a ~$\lambda^{\prime}-$type coupling, this could explain the ALEPH events.
A detailed analysis has not yet been performed.
Speculations~\cite{PECCEI} 
relating ALEPH events to the production of left handed squarks
are strongly constrained by the LEP1 precision data. New sources
of the custodial ~$SU_V(2)$ ~breaking in the left handed squark sector are
dangerous for the $\rho$--parameter and the related observables such as ~
$M_W$, ~$\sin^2\theta_{eff}^{lept}$ ~\etc. We have performed global fits to 
the electroweak data with a light ~($\sim55$ GeV) ~$\tilde b_L$ ~
for ~$m_{\tilde t_1}=115$ ~and ~$\tan\beta=2.4$ ~
(the case discussed in \cite{PECCEI}). While we agree with 
the authors in the evaluation of ~$\Delta\rho^{SUSY}(=0.00285)$,
the fit to ~11 ~electroweak observables
not involving heavy flavours (\ie\ ~$R_b$, ~$R_c$, ~$A_{FB}^{0b}$ ~and ~
$A_{FB}^{0c}$) gives ~$\chi^2\sim29$ ~which should be compared with ~
$\chi^2\sim7$ ~obtained for the case of a heavy ~$\tilde b_L$. ~

\begin{figure}[ht]
       \vskip 5in\relax\noindent\hskip -1.8in
       \relax{\includegraphics{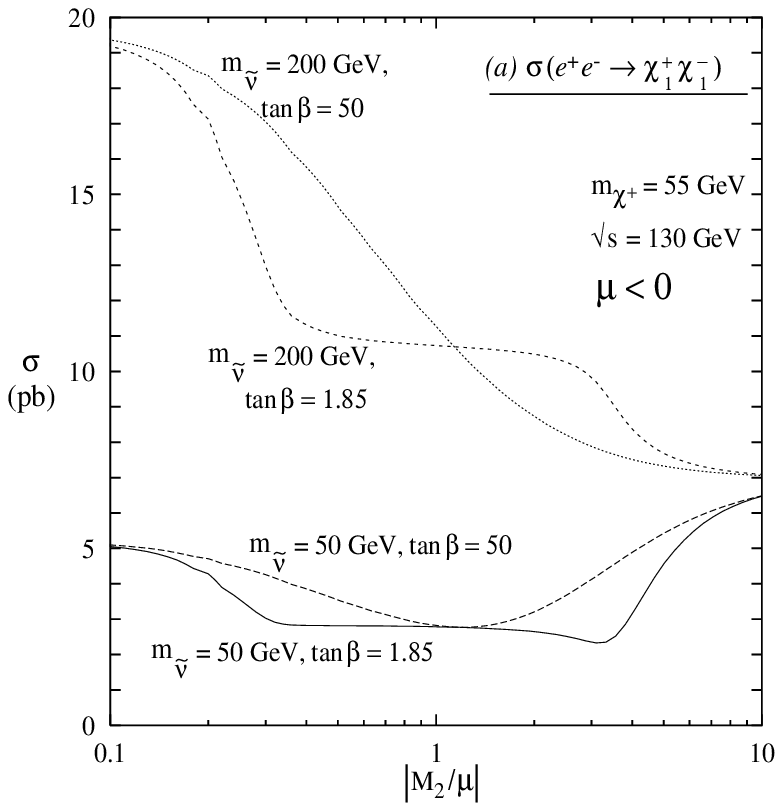}}
       \relax\noindent\hskip 3.25in
       \relax{\includegraphics{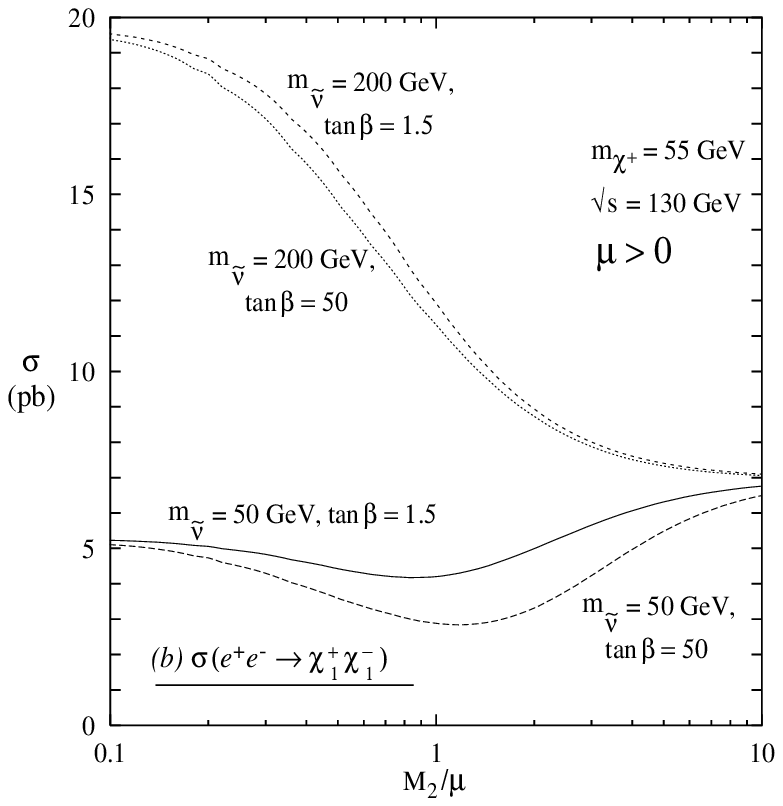}}
       \vspace{-20ex}
\caption{{\em Typical chargino pair production cross sections, at LEP1.5.
         {\em (a)} for $\mu < 0$
         {\em (b)} for $\mu > 0$.}}
\label{fig:gaugino_prod}
\end{figure}

Since the production cross sections for a neutralino of mass 55 GeV 
is too low, we are left with 
a light chargino  ~($m_{\chi^+}\sim 55$ GeV) ~as the most interesting 
candidate to explain ALEPH events. 
It survives both the total production cross section test 
(see Fig.\ref{fig:gaugino_prod}) and at the same time can explain 
the ~$R_b$ ~anomaly. It is, therefore, interesting to study the
effective chargino production cross section (with ALEPH cuts) and its
decay signatures in the \rpMS. In the rest of this analysis, we shall 
be conservative and confine ourselves to the ~$\mu > 0$ ~branch. The results 
for ~$\mu < 0$ ~are qauantitaively very similar (though it is more 
advantageous for ~$R_b$). ~We report
here  on a simulation at the parton level, followed by a brief comparison
with ALEPH observations.

If the chargino were to decay through a $L$--violating coupling, one should
see a relatively hard isolated 
lepton or have considerable missing momentum in the
event. Since such events have not been reported,
we shall here follow a different track, namely consider the 
$\lambda^{\prime\prime}-$type 
couplings\footnote{To the best of our knowledge, sneutrino pair production 
                   is the only
          possible scenario where a decay via the ~$\lambda^\prime-$type 
          couplings may have a role to play.}.

The chargino ~$\chi^+$ ~may then decay via either of two channels
(the asterisk symbolising possible offshellness)
\be
\chi^+\ra \tilde{q_1}^\ast q_2\ra q_2q_3 q_4 
\label{stopdecay}
\ee
or
\be
\barr{rl}
\chi^+\ra\Ne^\ast W^* \ra & \hspace*{-.5em}\Ne^\ast f_1 f_2 \\
                          & \hookrightarrow q_3 q_4 q_5 \,
\earr
\label{neutdecay}
\ee
with the neutralino `decaying' via an ~\rp\ mode. The chargino branching
ratios depend on the assumptions and are {\em a priori} unknown. For
example, if we have ~${\tilde q}_1 = {\tilde t}_R$ ~with ~
$m_{\chi^+} - m_b > m_{{\tilde t}_R}\gsim M_Z/2$, ~
the channel (\ref{stopdecay})
would tend to dominate ~($\gsim 90\%$), ~irrespective of the neutralino mass. 
On the other 
hand, if the squark was more massive than ~$\chi^+$ ~and the neutralino less,
the branching ratios would, in a large part, be determined by the
size of the relevant \rp\ coupling. It should be noted that a coupling like ~
$\lambda^{\prime\prime}_{tds}$ ~is rather weakly constrained \cite{BCS,BR}
and could still lead to a ~50\% ~branching fraction for the stop mediated
decay (even if the stop is significantly off-shell). 
The situation changes yet again if both the squark and ~$\Ne$ ~ 
are heavier than the ~$\chi^+$. ~Though such an eventuality does not
occur in a scenario with gaugino mass unification, it is instructive
to bear this possibility in mind. In view of the myriad possibilities,
we shall discuss the two cases separately assuming the corresponding
channel to be dominant.

As we are attempting only a parton level simulation (without incorporating
hadronization effects), it is not possible for us to mimic the exact
experimental 
cuts\footnote{Outstanding examples of such selection criteria are number 
              of charged tracks per jet, masses of individual jets \etc.},
but we shall try to be as close to the
experimental situation as possible. At the parton level we shall often 
encounter more than 4 quarks but not all these will lead to visible jets.
The reasons are manifold. Some of the jets may be soft or too close to the
beam axis and hence missed by the detector. To eliminate these, we impose
cuts on the energy and rapidity of jets:
\be
E({\rm jet}) > 0.1 \sqrt{s}, \qquad\eta({\rm jet}) < 3 \ .
\label{ghost-cut}
\ee
Any jet thus missed will contribute to the missing energy which is bound by
\be
E_{\rm missing}\lsim0.3\sqrt{s} \ .
\label{Esl-cut}
\ee
To distinguish jets, ALEPH imposes a Durham ~$y$-cut of ~0.008, \ie, for
any jet pair ~($ij$) ~to be distinguishable, they require
\be
y_{\rm Dur}\equiv2 \;{\rm min}(E_i^2, E_j^2) \ (1-\cos\theta_{ij}) > 0.008\ ,
\label{dur-cut}
\ee
where ~$E_i$ ~is the energy of the jet ~$i$ ~and ~$\theta_{ij}$ ~is the angle
between them. Any pair whose momenta do not satisfy the above condition
will then be merged to form a
single jet with momentum given by the sum of the constituent momenta, with
the process to be repeated 
iteratively\footnote{One might argue against applying such algorithms 
          to {\sl partons}
          and propose the cone algorithm (see eq.(\protect\ref{delta-cut})). 
          We have made the comparison and 
          while the efficiencies do change, the gross features of our 
          analysis remain unaltered.}.
An identical procedure is adopted with regards to a JADE ~$y$-cut:
\be
y_{\rm JADE}\equiv2 E_i E_j \ (1-\cos\theta_{ij}) > 0.022 \ .
\label{jade-cut}
\ee
To qualify, an event should have 4/5 distinguishable jets at this stage.
In case of 5 jets, the pair with the smallest invariant mass are to be 
merged to form a single jet. We are then left with 4 jets and hence 
6 ways to pair them. Each such pairing must satisfy
\be
m_{ij} > 25 \gev \ .
\label{minv-cut}
\ee
For the sample that has survived these cuts, jets are paired by selecting
the particular combination with the {\sl smallest} invariant mass difference
which has then to satisfy the condition
\be
\Delta m\equiv {\rm min}(m_{ij} - m_{kl}) < 20 \gev \ .
         \label{minv-diff-cut}
\ee
And finally, ALEPH finds that for this ``correct'' pairing, slightly more
than half ~(9/16) ~the events fall within the energy range
\begin{eqnarray}
103.3 ~{\rm GeV} < \Sigma m (\equiv m_{ij} + m_{kl}) < 106.6 ~{\rm GeV} \ .
\label{minv-sum-cut}
\end{eqnarray}
We shall, however, {\em not} use equation (\ref{minv-sum-cut})
as a cut. For one, the position of this bin might change with accumulation
of statistics or a change in the jet merging algotithms. Secondly,
it is interesting to see what kind of distributions are predicted 
within the scenarios that we are about to consider and compare these with 
the data. This is likely to afford us a better understanding of the 
operative physics.
\begin{figure}[ht]
       \vskip 5in\relax\noindent\hskip -1.8in
       \relax{\includegraphics{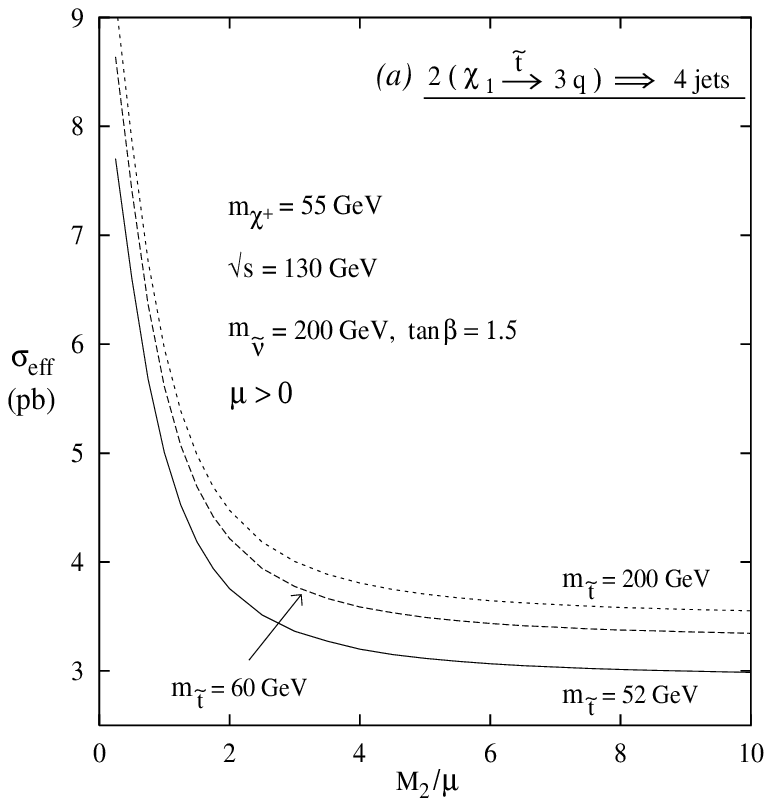}}
       \relax\noindent\hskip 3.25in
       \relax{\includegraphics{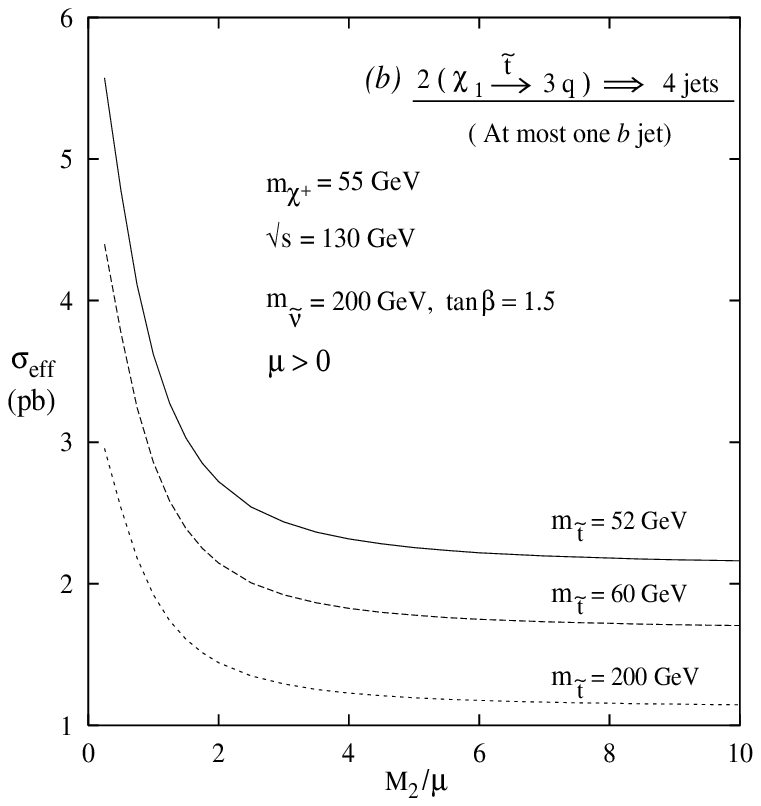}}
       \vspace{-20ex}
\caption{{\em {\em (a)} Typical effective (i.e. after the imposition of the 
         cuts of eqs.(\protect\ref{ghost-cut}--\protect\ref{minv-diff-cut}))
         $4-$jet cross section for pair--produced charginos 
         decaying solely through the stop channel 
         (see eqn.\protect\ref{stopdecay}).
         {\em (b)} As in {\em (a)}, but with the additional constraint that 
         at most one b-quark may contribute to the visible energy.} }
\label{fig:c_qs}
\end{figure}

If squark mediated decays (eqn. \ref{stopdecay}) were to be the dominant 
channel, the final state would comprise of six quarks, on which we must
impose the conditions described in 
eqns (\ref{ghost-cut}--\ref{minv-diff-cut}).
In Fig. \ref{fig:c_qs}{\em a}, we exhibit ~$\sigma_{\rm eff}$ ~
(for decays mediated by a virtual stop) as a function of the ratio
of the gaugino mass parameters ~$M_2/\mu$. ~While for small ~$M_2/\mu$, ~
we have relatively large cross sections, for large ~$M_2/\mu$, ~
we are quickly led to a 
plateau\footnote{We have not included here the small contribution from the 
                 direct production of a 52 GeV ~squark.}
of the order of the central value in eq.(\ref{aleph-cs}).
\begin{figure}[ht]
       \vskip 5in\relax\noindent\hskip -1.6in
       \relax{\includegraphics{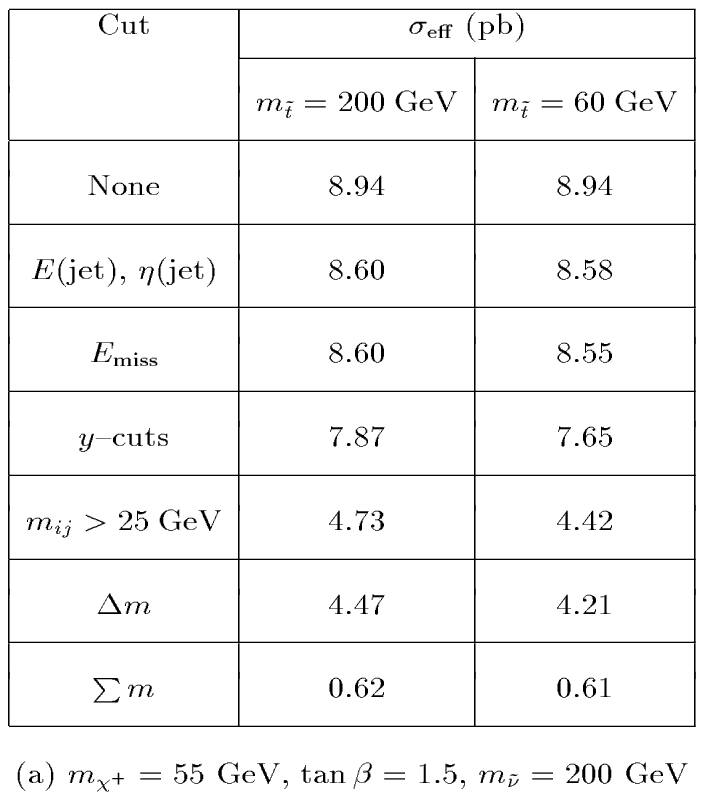}}
       \relax\noindent\hskip 3.15in
       \relax{\includegraphics{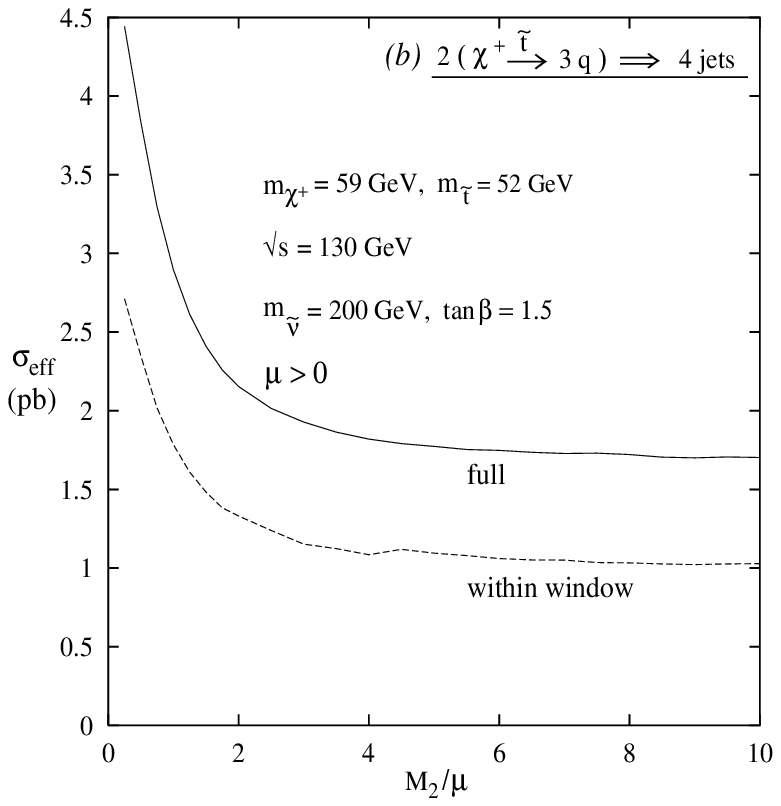}}
       \vspace{-20ex}
\caption{{\em {\em (a)}The effective cross-section, on imposition of various 
         cuts, for two typical values of the stop-mass. The efficiencies 
         are independent of $M_2 / \mu $ which has been chosen to be 2.0.
         {\em (b)}The effective cross-section for a 59 GeV chargino and 
         52 GeV stop. The lower(upper) curves are with(out) the cut of
         eqn.(\protect\ref{minv-sum-cut}). There are no $b$ events.}}
\label{fig:c59_s52}
\end{figure}

The {\em efficiency} obviously depends on the mass of the top-squark, and
the difference arises from different kinematical sources. The role of
the various effects can be seen from a study of
the table in Fig. \ref{fig:c59_s52}{\em a},
wherein we have tabulated the effective cross section after each of the
cuts mentioned in eqs.(\ref{ghost-cut}--\ref{minv-sum-cut}).

One may question
our explanation from the point of view that this particular decay mode
involves a $b$--quark. For example, if we do demand
that the observed jets carry {\em at most one $b$} (\ie, at least one $b$
leads to a ghost jet), we get a much smaller cross section
(see Fig.\ref{fig:c_qs}{\em b}). The difference thus constitutes the
sample where {\em both} the $b$--jets are visible (though often
merged with light quark jets). 
Thus, although ALEPH has a strong negative
result only for the ~$b\bar{b}b\bar{b}$ ~mode and the bounds on the
relative abundance of ~$b\bar{b}q_1\bar{q_2}$ ~channel are less precise, 
this still seems to constitute a potential problem for the scenario,
especially for larger ~$m_{\tilde t}$.

\begin{figure}[ht]
       \vskip 5in\relax\noindent\hskip -1.6in
       \relax{\includegraphics{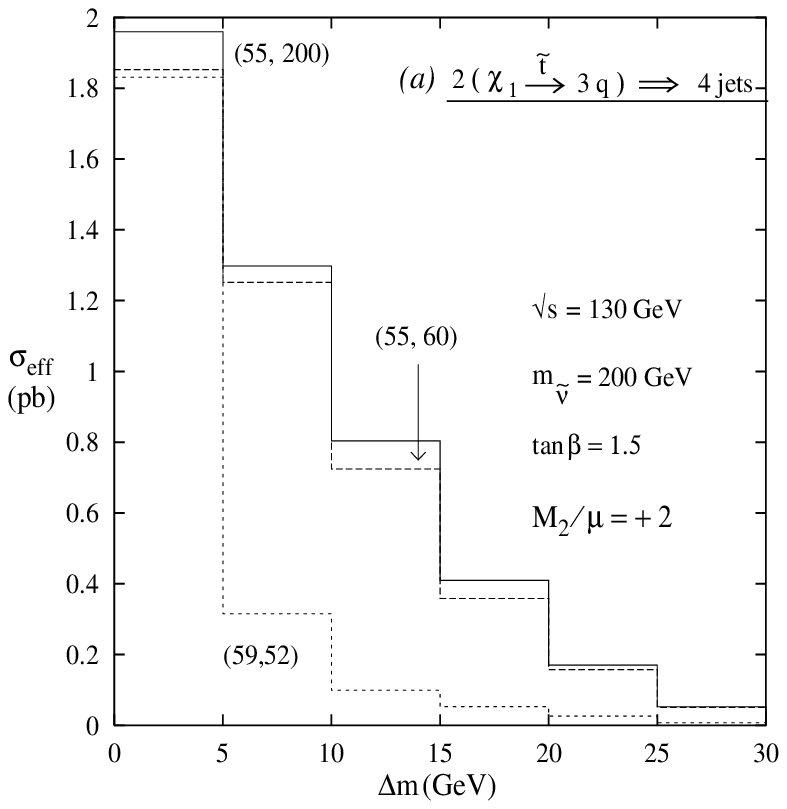}}
       \relax\noindent\hskip 3.15in
       \relax{\includegraphics{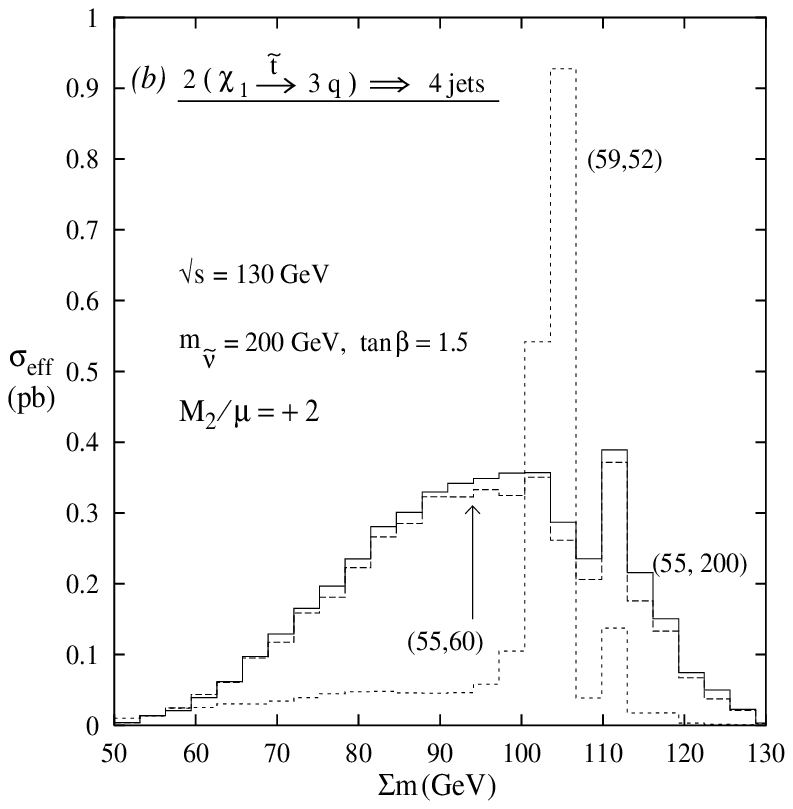}}
       \vspace{-20ex}
\caption{{\em {\em (a)}The variation of the effective cross-section
         with $\Delta m$  on imposition of the cuts of 
         eqs.(\protect\ref{ghost-cut}--\protect\ref{minv-cut}). The 
         numbers in the parentheses refer to the chargino and stop 
         masses respectively. 
         {\em (b)}As in {\em (a)}, but variation with $\sum m$ instead.
         }}
\label{fig:stop_distr}
\end{figure}

There is still another issue that we have not addressed as yet.
Does the \rp - induced 4-jet sample fall primarily within the 
$103.3$ GeV $<\Sigma m < 106.6$ GeV ~window? A look at the table in 
Fig.\ref{fig:c59_s52}{\em a} and Fig.\ref{fig:stop_distr}{\em b}
convinces us that this is not so. While
an efficiency factor of 55\% would have nearly mirrored the ALEPH 
data, we have a considerably lower fraction (about ~10-20\%) ~populating
this bin. The cause is not difficult to see. 
The chargino decay is a genuine three-body one leading to six quarks in 
the final state. Though these do merge, often it is not the right combination
that merge. Consequently, the invariant mass pairing would often choose the 
wrong combinations. Thus, this additional cut would shift the curves of 
Fig.\ref{fig:c_qs} much too low to explain the observed events.

Let us now explore the possibility that ~$m_{\tilde t}$ is small enough so
that the chargino can decay into an on-shell stop and a ~$b$-quark. Since
LEP already provides lower limits on ~$m_{\tilde t}$, ~we immediately see that
the ~$b$ ~quark will, almost always, 
be very soft. In fact, it would never survive the
cuts of eq.(\ref{ghost-cut}) and would be lost to the detectors. (In fact, 
a comparison of the $m_{\tilde t}$ dependence in 
Fig.\ref{fig:c_qs}{\em a} and Fig.\ref{fig:c_qs}{\em b} already 
points to this.)
All visible energy would now originate from the decay of the stops into
a pair of quarks each. Pairing mismatch effects are small, and hence 
most of the events could be expected to fall within the 
~$103.3 \gev < \sum m < 106.6$ GeV ~
window. Of course, to match with the ALEPH data, it is now ~
$m_{\tilde t}$ ~which is to be in the ~($51-54$) GeV ~range, with the 
chargino heavier by about ~$5$--$12$ GeV. In Fig.~\ref{fig:c59_s52}{\em b}, 
we exhibit the situation for a particular mass
combination \footnote{We have again neglected the direct
                      production of a stop-pair, as the corresponding total
                      cross section is small.}.
It should be noted that we have not folded in the actual 
$Br(\chi^+\ra{\tilde t}\;\bar b)$, ~but have assumed it to be ~
100\% ~(as stressed earlier, in the considered case this 
assumption is correct within about 10\% uncertainty).

For such a combination, both the ~$\Delta m$ ~and ~$\sum m$ ~
distributions (see Fig.\ref{fig:stop_distr})
turn out to be very close to the observed signal. While the table 
in Fig.\ref{fig:c59_s52}{\em a} already tells us an agreement 
in the ~$\Delta m$ ~is quite general, the $\sum m$ agreement 
is crucially dependent upon such a hierarchy. 
This, thus, offers a potentially simultaneous explanation of both 
ALEPH events and large value of ~$R_b$. ~
Such mass combinations are unique in our discussion in the sense of avoiding ~
$b$-jets as well as satisfying the distribution of (\ref{minv-sum-cut}). 

It is instructive to compare Fig.\ref{fig:c59_s52}{\em b} with the curves 
for ~$m_{\tilde t} = 52 \gev$ ~in Fig.\ref{fig:c_qs}. The crucial difference 
is the difference in the chargino and stop masses. In the former case,
the chargino clearly decays into a {\em real} stop, thus allowing the 
$b$-quark only a very limited phase space. As the mass difference shrinks,
the stop is pushed towards virtuality and more and more of the ~$b$s ~begin 
to acquire larger momenta. The point at which this effect takes over is 
of course determined by the stop decay width and hence on the magnitude of ~
$\lambda''_{tds}$.

Finally there is one more interesting remark to be made.
Had we considered a different squark above, the results
would have been similar. Moreover, we avoid problems with ~$b$ ~quarks
and can easily explain the magnitude of the cross section observed at 
ALEPH. Of course, there are also less satisfactory aspects now. For one, ~
${\tilde t}_R$ ~is expected to be the lightest squark both from the
point of view of the renormalization group evolution as well as
considerations of ~$R_b$. ~Also, ~\rp\ couplings such as ~
$\lambda^{\prime\prime}_{tds}$ ~are very weakly constrained and thus may 
be large enough for the chargino decay to be dominated by stop exchange. 
Nor should it be forgotten that requiring a 
sharp ~$\sum m$ ~distribution would then necessitate the squark 
to be almost degenerate with the chargino (since the quark at the primary
decay vertex is now very light). 

\begin{figure}[h]
       \vskip 5in\relax\noindent\relax\hskip -1.8in
       {\includegraphics{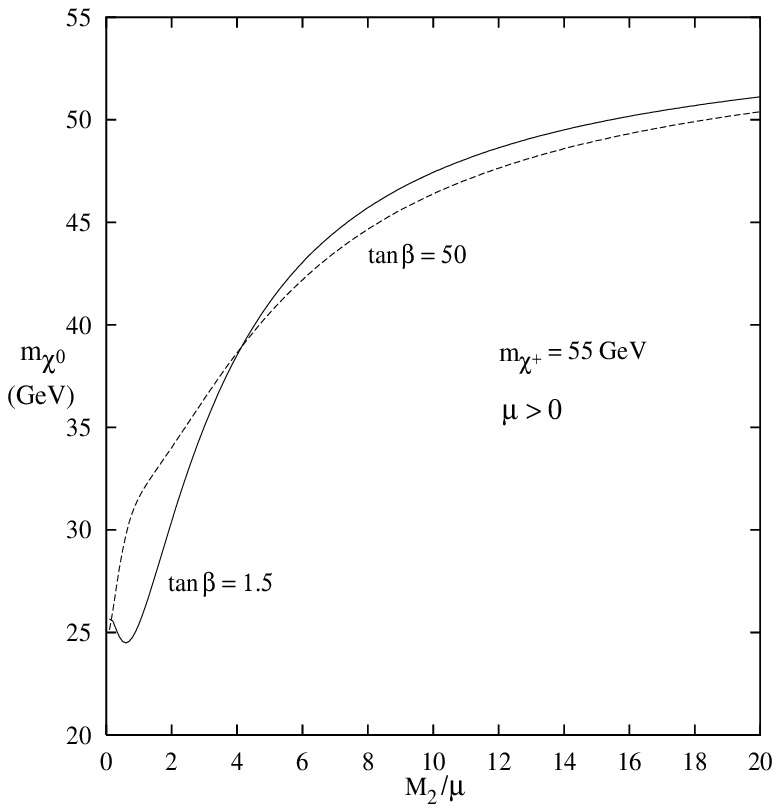}}
       \relax\noindent\hskip 3.25in
       \relax{\includegraphics{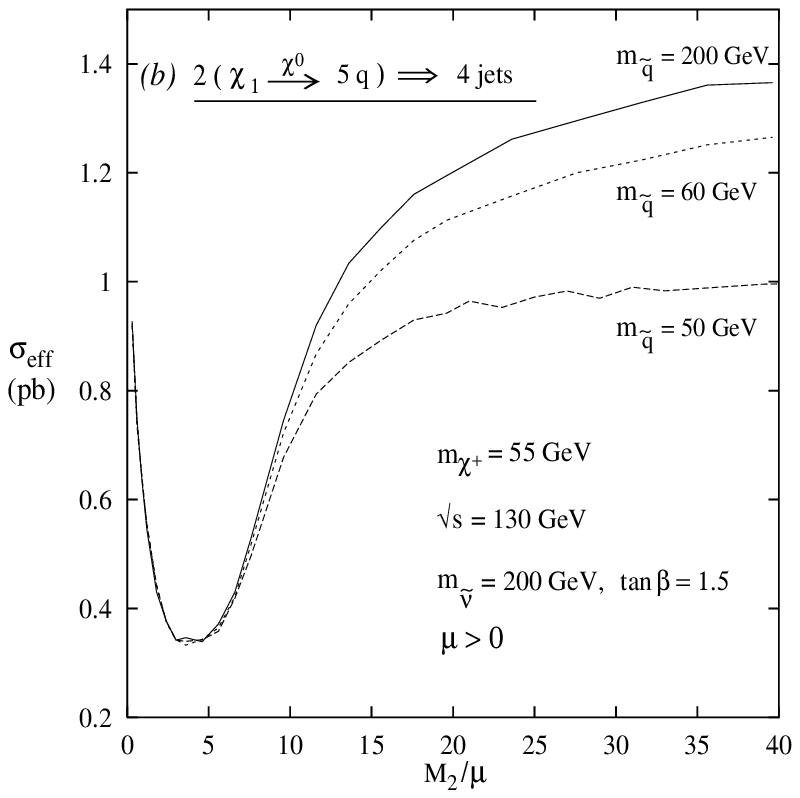}}
       \vspace{-20ex}
\caption{\em {\em (a)} The lightest neutralino mass as a function of
         $M_2/\mu$ for fixed chargino mass of 55 GeV.
         Gaugino mass unification has been assumed.
         {\em (b)}Typical effective ({\em i.e.} after the imposition of the 
         cuts of eqns.(\protect\ref{ghost-cut}--\protect\ref{minv-diff-cut}))
         $4-$jet cross section for pair--produced charginos
         decaying solely through the neutralino channel
         (see eqn.\protect\ref{neutdecay})}.
\label{fig:neutmass}
\end{figure}

We now move on to the decay channel of eq. (\ref{neutdecay}). At this point, 
two comments are in order:
\begin{itemize}
\item We assume that ~$\Ne$ ~is lighter than ~$\chi^+$. 
      To be specific, we make the assumption of universal 
      gaugino masses at the unification scale. The latter would thus prefer to 
      decay into an on-shell ~$\Ne$ ~and a pair of fermions. Since the mass 
      splitting between ~$\chi^+$ ~and ~$\Ne$ ~is a function of ~$M_2/ \mu$ ~
      (see Fig. \ref{fig:neutmass}{\em a}), one envisions an additional 
      dependence on this ratio (on account of the kinematical distribution of 
      the decay particles) over and above the dependence through the 
      production cross section.
\item If $\lambda^{\prime\prime}_{tds}$ were the only
      non-zero ~\rp\ coupling, ~$\Ne$ ~would have a ~5 ~body decay rather than
      a ~3 ~body one. The final state will thus have ~14 ~particles in it. To
      simplify the discussion as well as to keep our simulations at a tractable
      level, we rather consider a scenario where some other coupling
      (such as ~$\lambda^{\prime\prime}_{cds}$) ~is non-zero instead and 
      one of the associated squarks is the lightest. This 
      choice is somewhat 
      unaesthetic\footnote{Strictly speaking, in the presence of the baryon
                    number violating couplings, the lower mass bound on 
                    right handed squarks of the first two generations are 
                    around 45 GeV. Nevertheless, as mentioned eralier, one 
                    expects the stop to be the lightest squark.}
      but, hopefully, gives us qualitatively similar results to the 
      ~$5-$body decay.
\end{itemize}

The final state can then be one of three: ~$10 q$, ~$8 q + l \nu$ ~and ~
$6 q + l l' \nu \nu'$ ~with relative fractions close to ~$4:4:1$. ~We consider
first the purely hadronic decay channel.

The large event multiplicity, at the parton level, may lead one to
suspect that most of the events would actually lead to more than ~5 ~jets.
However, the very multiplicity leads to many of the quarks being either
soft or close to each other. The same combination of phase space
cuts and jet merging algorithm, as discussed earlier, then forces a 
large fraction of the possible final states to merge into ~4/5 ~jets.

In Fig.\ref{fig:neutmass}{\em b}, we display the effective cross section
for this mode as a function of ~$M_2/\mu$. ~The effective cross sections
are significantly lower than those obtained for the squark mediated decay. 
Even more striking is the difference in the ~$M_2 / \mu$ ~dependence. Comparing
these curves with that of Fig.\ref{fig:gaugino_prod}$b$, we see that the
efficiency is the least for comparatively lower values of ~$M_2/\mu$. ~The
reason is not difficult to understand. For smaller values of this ratio, the
neutralino is significantly lighter than the chargino
(see Fig.\ref{fig:neutmass}{\em a}) and thus the quarks from the primary decay
vertex are, on the average, quite energetic. Drawing on the experience
of the stop--mediated decay, one immediately envisages the rather large
probability of wrong merging and pairing of jets. 
This feature is likely to survive for ~5-body decays of the neutralino as 
well.

Accomodating ~$M_2/\mu \lsim 25$ ~(in a scenario with gaugino mass 
unification) is difficult on two accounts. A
small value for this leads to configurations
where all the quarks are quite energetic. Consequently, a larger
fraction of events end up with {\em more than ~5 ~jets}.
As ALEPH has not reported events with such large multiplicity,
the scenario stands on a weak ground. Also,
note that ~$\Ch \ra \Ne f f'$ ~decay has only ~67\% ~hadronic branching
fraction. Thus we should have expected to see quite a few events
with ~3 ~or ~4 ~jets and an isolated energetic lepton (possibly with
significant missing momentum as well). We have simulated such
decay modes with the requirement that the isolated lepton 
carries a minimum of ~10 GeV. ~To quantify isolation, we use
\be
\Delta R \equiv \left[ (\Delta \phi)^2 + (\Delta \eta)^2\right]^{1/2}
\label{delta-cut}
\ee
where ~$\Delta\phi$ ~and ~$\Delta\eta$ ~represent respectively
the azimuthal separation
between the lepton and a jet and the difference in their rapidities.
Choosing ~$\Delta R = 0.5$, ~we find that, for the parameter space
of interest, at most ~3 ~events with ~(3 jets + isolated lepton) and ~
2 ~events with ~(4 jets + isolated lepton) are expected. The SM
backgrounds arise from both QCD processes and ~$b \bar{b}$ ~production.
No such excess have been reported as yet.
While it may be still too early to rule out this possibility, it
would, perhaps, be better if a proposed scenario does not lead to
such events. This can be ensured only 
the chargino and the neutralino are closely 
degenerate\footnote{This requires either a very large ~$M_2/\mu$ ~
                    (as in our case)
                    or a relaxation of the gaugino mass unification 
                    assumption (the latter, as dicussed earlier, 
                    also leads to a better fit to the precision data).},
when  the leptons (and the neutrinos) would
again be very soft and hence not meet the selection criteria.

\begin{figure}[htb]
       \vskip 5in\relax\noindent\hskip -1.7in
       \relax{\includegraphics{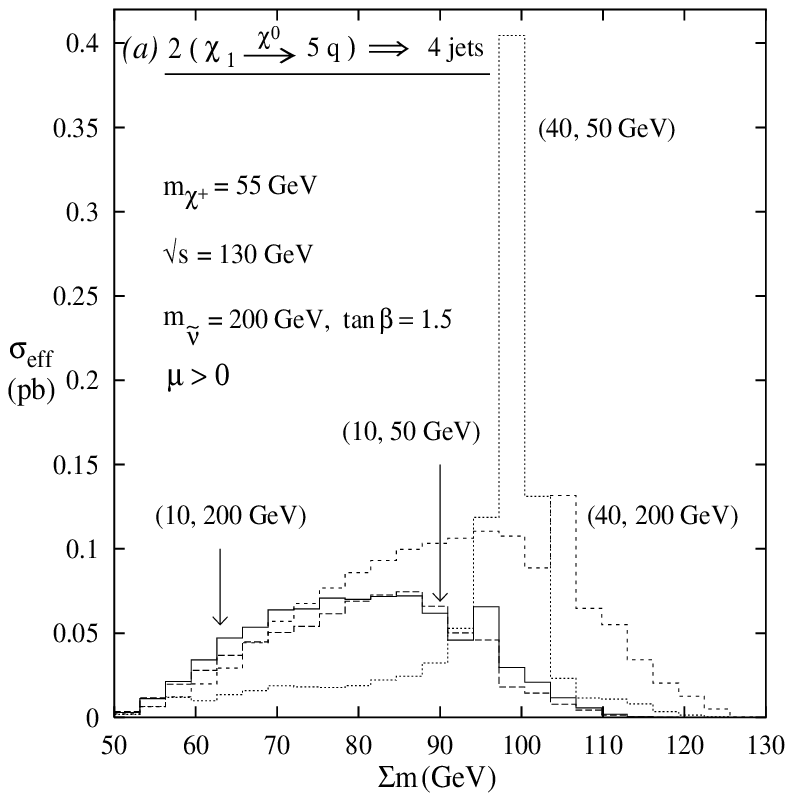}}
       \relax\noindent\hskip 3.3in
       \relax{\includegraphics{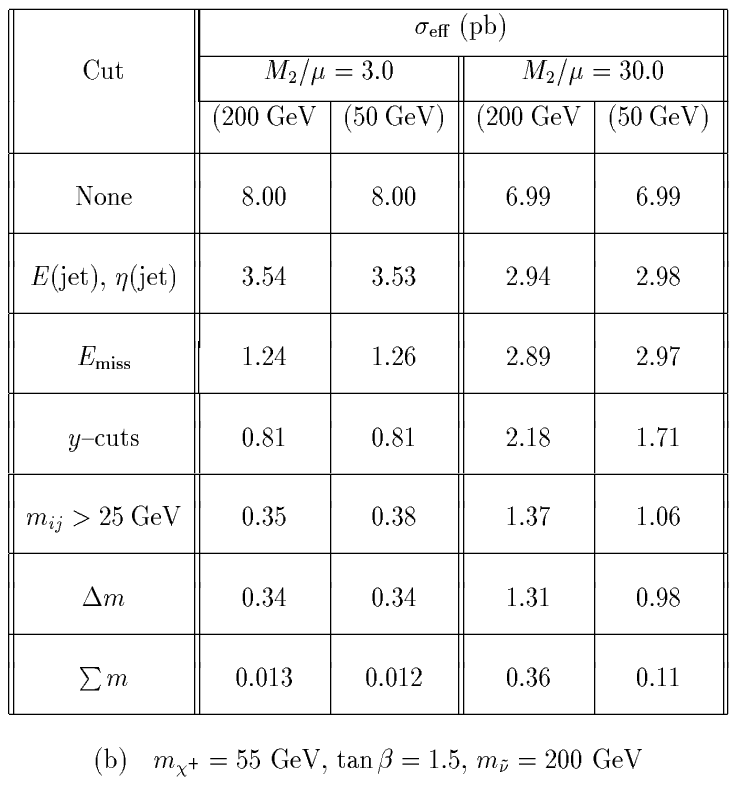}}
       \vspace{-20ex}
\caption{\em {\em (a)}The $\sum m$ dependence of the effective cross section,
         for various ($M_2/\mu$, $m_{\tilde q}$) pairs. All cuts except that 
         of eqn. (\protect\ref{minv-sum-cut}) have been imposed.
         {\em (b)} As in Fig.\protect\ref{fig:c59_s52}{\em a}, but for 
         the neutralino--dominated decay channel. The numbers in 
         the parentheses represent the relevant squark mass.}
\label{fig:c_nw_distr}
\end{figure}
A close degeneracy is ``useful'' on another account too. As
in heavy stop mediated decays, here too we expect the mass
distribution to be relatively broad. That this is indeed the case is
borne out by a look at Fig.\ref{fig:c_nw_distr}{\em a}. Only 
when the chargino and the neutralino are
almost degenerate, and hence the primary decay quarks are  soft to be
observed, can we have a relatively sharp distribution. Here too a
distinction may be made between the heavy squark and the light squark
case. Since a large ~$M_2/\mu$ ~essentially means that the chargino serves
only to increase the neutralino production cross section and that we may
instead focus on the decays of the latter, for such a scenario we can
borrow the results of the previous section with the chargino replaced by
the neutralino. Thus for a squark just about lighter than the
neutralino\footnote{Though in this case, the chargino will tend to
decay directly into this squark (unless it is a ~$\tilde b$), ~we neglect
this effect for the time being.},
the quark from the primary (neutralino) decay vertex would be soft
(unobservable) and thus effectively we return to squark pair-production
(albeit with enhanced cross section), with each decaying subsequently
to a pair of quarks. The squark mass dependence can be inferred from the 
table in Fig.\ref{fig:c_nw_distr}{\em b}.

We may then safely infer that if the ~$\chi^+_1\ra\Ne ff^{\prime}$ ~decay 
mode is to be the dominant one, then a satisfactory explanation
of the ALEPH events necessitates both a close degeneracy between the 
chargino and neutralino masses, and 
a squark just about lighter than the neutralino. Clearly, a 
scenario with the chargino decaying directly into a such a light 
squark is more attractive.

To conclude, we
point out that broken ~$R$--parity (specifically the existence
of the baryon number--violating ~$\lambda''$ ~couplings) could provide
an explanation for the excess four-jet events observed by the
ALEPH collaboration. Though several mechanisms have already been
proposed~\cite{FARRAR,PECCEI}, the scenarios therein are in
considerable disagreement with the global fits. We instead focus on
a light chargino, which subsequently decays into ~3 ~(stop-mediated)
or ~5 ~(neutralino mediated) quarks each. Parton level simulation
show that, for a considerable fraction of the events, the resultant
jets merge to give a signal similar to that observed by ALEPH. While
this is true for a large region of the parameter space, a close
agreement with the observed (albeit with low statistics)
invariant mass distribution is possible
only for a more restrictive set. The best fits are obtained for a
chargino decaying predominantly into a stop-bottom pair, with the extremely
soft $b$ escaping detection and the stop ~($\sim 52$ GeV) ~decaying into
a pair of quarks. It is interesting to note that, for such a
choice of parameters, ~$R_b$ ~can indeed be significantly larger
than the SM value.

We must reiterate that our analysis is only a parton level one,
and hence our results are at best indicative. A more definitive
statement would require a proper simulation including, amongst
others, hadronization effects and detector efficiencies. 
What lends additional credence to our parton level analysis is a 
comparison (for some of the stop-mediated decay modes) with 
a full Monte Carlo analysis with quark fragmentation in JETSET and
a simulation of the ALEPH detector. The simple parton-level and 
the complete
analyses agree at the ~15\% ~level in both the accepted number of 
events and the shape~\cite{CARR}.

{\bf Acknowledgements}: We are grateful to John Carr 
for checking our results with a full detector level Monte Carlo
simulation and for his interest in this analysis.
DC thanks S.~Raychaudhuri and R.~Settles for
many useful discussions. 
The work of PHC was partly supported by 
the joint U.S.-Polish Maria Sk\l odowska Curie grant.
 Part of the work was done during DC's visit
to the Institute of Theoretical Physics, Warsaw University and he
expresses his thanks for their support.

{\em Note added}: The possibility that the 4-jet events can be
interpreted in terms of pair-produced charginos (and neutralinos) decaying
through the the $\lambda''$ couplings has also been considered
in ref.\cite{GGR}. The authors agree that neutralino production
cross sections are too small, and that an explanation in terms of
charginos decaying through the neutralino channel, while viable, is
tenuous. However, they have not examined the (better) decay mode of
eq.(\ref{stopdecay}). Also, the connection with $R_b$ has not been
explored.

\end{document}